\documentstyle[12pt]{article}
\newcommand{\newc}{\newcommand}
\newc{\beq}    {\begin{equation}}
\newc{\eeq}    {\end{equation}}
\newc{\beqa}    {\begin{eqnarray}}
\newc{\eeqa}    {\end{eqnarray}}
\newc{\st}    {\stackrel}

\begin{document}

\renewcommand{\baselinestretch}{2}
\large
\normalsize
\noindent

\begin{titlepage}
\title{ Galactic Halos as Boson Stars}
\author{ Jae-weon Lee
 and In-gyu Koh  \\ \it Department of Physics,
 \\ \it Korea Advanced Institute of Science and Technology,
\\   \it 373-1, Kusung-dong, Yusung-ku, Taejeon, Korea  \\
\\
}

\date{}
\maketitle
\vspace{-9cm}
\vspace{10cm}

We investigate the boson star
with the self-interacting scalar field as a
model of galactic halos.
 The model has slightly increasing  rotation curves
and allows wider ranges of the mass($m$) and
coupling($\lambda$)
of the halo dark matter particle than
 the  non-interacting model previously
suggested(ref.\cite{sin1}).
Two quantities are related by $\lambda^{\frac{1}{2}}
(m_p/m)^2\st{>}{\sim} 10^{50}$.


\vspace{4cm}
\maketitle

\end{titlepage}
\newpage

\vskip 5mm

It is well known that the flatness of the  galactic
rotation curves indicates the presence of dark matter
around galactic halos.
However, the properties of the dark matter are
still  mysterious.
For example, why the dark matter in halos does not fall
towards the center of galaxy and form black holes?
The answer to the above question may be
a good  criterion for the good halo model.

There are thermal distribution model \cite{thermal} where
density profile $\rho\sim r^{-2}$
, and   spherical infall model\cite{infall}
where $\rho\sim r^{-2.25}$.

Recently  Sin\cite{sin1,sin2} suggested a new model
of the halos composed of pseudo Nambu-Goldstone boson
(PNGB).

According to the model, the condensation of ultra
 light PNGB
 whose Compton wavelength
$\lambda_{comp} = \frac{\hbar}{mc}$
is about $R_{halo}$
is responsible for the halo formation.
Cosmological role of the ultra light PNGB
was studied
in the late time phase transition model
\cite{late} to reconcile the smoothness in
the background radiation
with the large scale structure.

Before Sin's work,  an astronomical object
 which consists of
the PNGB dark matter was suggested by some
authors\cite{dark}.
In their model the force against gravitational
 collapse comes from the momentum uncertainty
of the quantum mechanical uncertainty principle.

 Since the typical length scale $R$ in this model
 is  Compton wavelength $\lambda_{comp}
  \sim \frac{1}{m}$ of the particle,
  the typical mass scale of the object
 is $M\sim \frac{R}{G} \sim \frac{m^2_p}{m}$.

Similarly, in Sin's model  galactic halos
are the objects of the self-gravitating bose
 liquid whose collapse are
prevented by the uncertainty principle.

The typical halo has radius
$R_{halo}\sim 100kpc\sim 10^{24}cm$
and mass $M_{halo}\sim 10^{12} M_{\odot}\sim 10^{45}g$,
so one find the mass $m$ of the PNGB  whose de Broglie
wave length $\sim R_{halo}$
is about $10^{-26} eV$.

Note that the de Broglie length $\sim \frac{c}{v} \times \lambda_{comp}$
is more adequate  to our purpose.

The self-gravitating condensed states
 are described by the following
non-linear Schr\"oedinger equation:
\beq
i\hbar\partial_t \psi =-\frac{\hbar^2}{2m}\nabla^2\psi+
GmM_0 \int^{r'}_0 dr'\frac{1}{r'^2}\int^r_0 dr'' 4\pi r''^2
 |\psi|^2\psi(r),
\label{sch}
\eeq
which was known as the Newtonian limit of the boson star
fields equation\cite{newt1}.

The normalization constant $M_0$ is chosen to
give the total mass of halo  $M=M_0\int dr 4\pi r^2|\psi|^2$
as in ref.\cite{sin1}.

The rotation velocity of the stellar object rounding halo
at radius $r$ is given by
\beq
 V(r)=\sqrt{\frac{GM(r)}{r}},
\label{vr}
\eeq
where $M(r)$ is mass within $r$.

Integrating eq.(\ref{sch}) numerically and using
eq.(\ref{vr}) Sin found slightly increasing
rotation curves and density profile $\rho\sim r^{-1.6}$.

What happens if there are repulsive self-interactions
between the dark matter particles?
To answer this and stability question it is desirable
to study the relativistic fields equations than
the Schr\"oedinger equation.

The cold gravitational equilibrium configurations
of massive scalar field were found by solving
the Klein-Gordon equations with gravity
decades ago\cite{orig}.
 We find that these configurations, called boson star
\cite{nontopo}, are adequate to the relativistic
 extension of Sin's model.

Consider a self-interacting complex scalar field
 and the gravity whose action is given by
\beq
 S=\int \sqrt{-g} d^4x[\frac{-R}{16\pi G}
-\frac{g^{\mu\nu}} {2} \phi^*_{;\mu}\phi_{;\nu}
- \frac{m^2}{2}|\phi|^2 -\frac{\lambda}{4}|\phi|^4].
\eeq
Since halos seem to be spherical, we choose
Schwarzschild metric
\beq
 ds^2=-B(r)dt^2+A(r)dr^2+r^2 d\Omega
\eeq
and assume spherically symmetric field solutions
\beq
\phi(r,t)=(4\pi G)^{-\frac{1}{2}} \sigma(r)
e^{-i\omega t}.
\eeq
{}From the action, dimensionless time independent Einstein and
scalar wave equations appear as in ref.\cite{colpi}:
\beq
\frac{A'}{A^2x}+\frac{1}{x^2}[1-\frac{1}{A}]=[\frac
{\Omega^2}{B}+1]\sigma^2+
\frac{\Lambda}{2}\sigma^4 +\frac{\sigma'^2}{A},\\
\eeq
\beq
\frac{B'}{ABx}-\frac{1}{x^2}[1-\frac{1}{A}]=[\frac
{\Omega^2}{B}-1]\sigma^2-
\frac{\Lambda}{2}\sigma^4 +\frac{\sigma'^2}{A},\\
\eeq
\beq
\sigma''+[\frac{2}{x}+\frac{B'}{2B}-\frac{A'}{2A}]\sigma'
+A[(\frac{\Omega^2}{B}-1)\sigma -\Lambda \sigma^3]=0,
\label{eqs}
\eeq
where $x=mr, \Omega=\frac{\omega}{m}$ ,
$A\equiv[1-2\frac{M(x)}{x}]^{-1}$ and $\Lambda=
\frac{\lambda m^2_p}{4\pi m^2}
$.

One may take $M(x)$ for dimensionless mass of the
boson star for large $x$.

Numerical solutions of the above
equations are studied by many authors\cite{numerical,free,density}.  The
required boundary conditions are $~M(0)=0,\sigma'(0)=0$ and $ B(\infty)=1$ and
free parameters are $\sigma(0)$ and $\Omega$.

For the case  $\Lambda=0$ \cite{free} it was found that
there is a maximum  mass
 $ M_{max}=0.633\frac{m^2_p}{m}$
for the zero node solution.

We will focus
on the non-zero node solutions,
because the rotation
curve of the zero node solution falls too fast
to explain the flatness of the rotation curves
of many galaxies(see ref.\cite{sin2} for more
arguments).
This raises the stability problem of higher node
solutions, which will be discussed later.

Maximum masses  for higher node solutions are
 proportional to node number $n$ and
about the same order  as for the zero node
case for small $n$.

This with $M_{halo}$ gives us
$ m \st{<}{\sim} 10^{-22} eV$.

Another constraint comes from
the maximum stable center density
against small radial perturbation  \cite{density}
 $\rho_{c}=2.1\times 10^{98}
 m^2 g/cm^3 > 10^{-24} g/cm^3$ ,which is equivalent
 to $ m \st{>}{\sim} 10^{-28} eV$ for the
 zero node solutions.
So for the zero node solutions
   $10^{-28}eV \st{<}{\sim} m \st{<}{\sim} 10^{-22}eV$.

For the case  $\Lambda\neq 0$, new scale appears because of the
  repulsive force preventing halo from gravitational
 collapse.
In this case the typical length scale
 is $R\sim \Lambda^{\frac{1}{2}} /m$,
thus the typical mass scale is $\frac{R}{G}
\sim \Lambda^{\frac{1}{2}} m_p^2/m$,
which is also of order maximum mass
like $\Lambda=0$ case.

Numerical study\cite{colpi} shows
 $M_{max}=0.22 \Lambda^{\frac{1}{2}}\frac{m^2_p}{m}$
for zero node solutions.
 From the fact that $M_{max}>M_{halo}$ we find
\beq
\lambda^{\frac{1}{2}}(\frac{m_p}{m})^2\st{>}{\sim} 10^{50}.
\label{mlambda}
\eeq
This is a relation between the mass and  coupling of
 the halo dark matter particle.
For the perturbative case ( $\lambda<O(1)$) the above
relation implies $m\st{<}{\sim}10^3eV$.

To treat particles as a classical field, we
require that the  inter-particle
distance should be smaller than their Compton wave length.
 This gives $m\st{<}{\sim}10^{-2} eV$.

Note that
 $\Lambda= \lambda m_p^2/4\pi m^2$
is very large even for very small $\lambda$
 due to the smallness of $m$
 relative to $m_p$
,hence the
self-interaction effect is non-negligible.

Were there any realistic particle physics model
satisfying the above relation?
Unfortunately, the usual cosine potential $V(\phi)=\mu^4(1-Cos(\phi/f))$
 for the PNGB is inappropriate for our study,
because the sign of the quartic coupling constant is negative
in the Taylor expansion about the potential  minima
 and $\phi$ is real.
Real scalar field such as axion may form oscillating
soliton star\cite{soliton}
rather than the boson star.

Instead, we consider the following potential.
\beq
V(\phi)=\mu^4(1+(\frac{\phi}{f})^2)^2.
\eeq
Inserting mass and quartic coupling from
the above potential to
the relation in eq.(\ref{mlambda}), we get
$0.1(\frac{m_p}{\mu})^2\st{>}{\sim} 10^{50}$
 and equivalently $\mu\st{<}{\sim} 10^2eV$.

We also solve the equations numerically and find
 the dimensionless rotation velocity which is given by
$V_{rot}=\sqrt{M(x)/x}= [\frac{1}{2}(1-A^{-1})]^{\frac{1}{2}}$.

The results are shown in fig.1 and fig.2.

Fig.1 shows rotation velocity curves
for the cases $\Lambda=0$ and $\Lambda=300$ .
The parameters are $B(0)=0.641 , \sigma(0)=0.1$
and $B(0)=0.781,\sigma(0)=0.01$ respectively.

Fig.2 shows $\sigma$ and rotation velocity curve of
8 nodes solution($n=9$).

It is interesting that the line connecting
 minimum points of the rotation velocity is almost straight.
For large $n$ and  $\Lambda \gg 1$
 mass profile is $\rho\sim r^{-1.7}$.
Rotation curves are slightly increasing
regardless of the self-interactions.

Including the visible matter may change the slope
of the curves  and  explain the variety of
the observed galaxy rotation curves as shown
in ref.\cite{sin2}.

We will now study the Newtonian limit of our model.
The strength of the gravity of halo $G M_{halo}/R_{halo} $ is
comparable to that of the earth.
  Therefore we can use
the Newtonian limit $\Omega=\frac{\omega}
 {m}=\sqrt{1+(\frac{k}{m})^2}\rightarrow 1$
 ,which is comparable to the Newtonian gravity
approximation $2M(x)/x \ll 1$.
Collecting terms to $O(\xi^2=(\frac{k}{m})^2)$
one can find that for $\Lambda=0$ the equations
 of motion are\cite{newlimit}
\beqa
\nabla^2 \sigma = \gamma\sigma,
\label{sigma}\\
\nabla^2 \gamma = 2\sigma^2,
\label{gamma}
\eeqa
where $\gamma\equiv 1-\frac{\Omega^2}{B}$.

Integrating eq.(\ref{gamma}) and inserting the
 result into eq.(\ref{sigma}) we can find
\beq
\frac{1}{2}\nabla^2 \sigma = (E+\int^x_0 dx^{'} \frac{1}{x^{'2}} \int^{x'}_0
dx^{''}x^{''2} \sigma^2)\sigma,
\eeq
which is  a dimensionless version of eq.(\ref{sch}).
Here $E$ is a integral constant.

Therefore one may treat the Bose liquid model as
a Boson star model.

 It is useful to study the scaling
 properties of eq.(\ref{sigma}) and eq.(\ref{gamma})
 for analyzing numerical solutions.

 Rescaling the total number of charges $N=Q\propto
\int\sigma^2 x^2dx$ $l$ times
increases the mass $l$ times.

 Equations (\ref{sigma}) and (\ref{gamma})
 are invariant under this rescaling when
\beq
x\rightarrow l^{-1} x, \sigma \rightarrow l^2 \sigma,
 \gamma \rightarrow\l^2\gamma,
\label{scal1}
\eeq
which is consistent with the model in ref.\cite{sin1}.

So one may say that for the non-interacting case the heavier halos
are smaller in size.

 It is difficult to find the scaling
properties for the case $\Lambda\neq0$.

For the case $\Lambda\gg1$ further rescaling :
$\sigma_*=\sigma\Lambda^{1/2}, x_*=x\Lambda^{-1/2}$ and $
M_*=M\Lambda^{-1/2}$
with neglecting terms to $O(\Lambda^{-1})$
yields  wave equations:
\beq
\sigma_*^2=(\Omega^2/B-1)=-\gamma,
\eeq
\beq
M'_*=\frac{1}{4} x_*^2(3\Omega^2/B+1)(\Omega^2/B-1),
\label{m}
\eeq
\beq
\frac{B^{'}}{ABx_*}-\frac{1}{x_*^2}(1-A^{-1})
=\frac{1}{2}(\Omega^2/B-1)^2,
\label{b}
\eeq
which are also shown in ref.\cite{colpi}.

Following the same procedure of $\Lambda=0$ case, we
get the Newtonian limits of eq.(\ref{m}) and eq.(\ref{b}).

\beq
\nabla^2\gamma=2\sigma_*^2=-2\gamma,
\eeq
whose solutions are
\beq
\gamma=-\gamma_0\frac{Sin(\sqrt{2}x_*)}{\sqrt{2}x_*},
\eeq
and
\beq
\sigma_*=\sqrt{\frac{\gamma_0 Sin(\sqrt{2}x_*)}{\sqrt{2}x_*}},
\eeq
where $\gamma_0=|\gamma(0)|$.

The above approximation is invalid when $x_*$ is large
and $n>1$.

As expected, the typical length scale is
 $ m^{-1} \Lambda^{\frac{1}{2}}$.
These solution do not show simple scaling property, however
numerical study indicates that when the central density
is lesser than the critical value corresponding to $M_{max}$,
heavier halo has smaller radius
 for both $\Lambda =0$ and $\Lambda\neq0$ cases.
\cite{numerical}

Note  the facts that the above arguments are valid when
the node number is fixed and
both mass and radius of boson star are increasing function of
the node number.

Now, let us discuss the stability of higher node solutions.
There are studies\cite{free,density} indicating
that non-zero node
solutions with $\Lambda=0$ are
unstable against fission and the small
radial perturbation.

Since higher node solutions are unstable,
they  must be long-lived to explain the age of
galaxies.

One possible decay mechanism for non-zero
node solution to zero node solution
  is gravitaional radiation\cite{gw}.
The power of gravitaional radiation $P$ is
about $G(d^3 I/dt^3)^2$, where the quadrapole
moment $I\sim M_{halo} R_{halo}^2$.
The available time scale is
$T\sim(R_{halo}^3/GM_{halo})^{1/2}$,
which is given by the virial theorem.
The parameter $\alpha\equiv GM_{halo}/R_{halo}$
indicates how much halo is relativistic.

So we can find a crude estimate of power
$P\sim G^4(M_{halo}/R_{halo})^5\sim\alpha^5/G
\sim \alpha^5 10^{59} ergs/s$.
Since $\alpha\sim 10^{-7}$ for halo,
$P\sim 10^{24} erg/s$.
The potential energy of halo $GM_{halo}^2/R_{halo}$
is about $10^{58}erg$, and  therefore time scale
of the decay by the gravitational radiation
   is much longer than the age of galaxies
 $\sim 10^{10} yr \sim 10^{17} s$.

For the $\Lambda\neq0$ case, there is a work
indicating that higher node solutions are stable
against the perturbation with fixed particle number
\cite{nfix}.
 However, there seems to be no work on the
stability against the more general perturbations.
Since it is still unclear that higher node solution with
$\Lambda\neq0$ is stable, we must again
estimate the life time of halo against gravitational
radiation.
{}From eq.(15) we get $\Lambda \sigma^4=-\gamma \sigma^2$
, which indicates that in Newtonian limit
the energy of the repulsive force
is comparable to the gravitaional potential energy.
So the energy distribution in halo
is not so different from the $\Lambda=0$ case.
Therefore, we argue that the same procedure for
calculating the gravitaional radiation
is applicable to the $\Lambda\neq=0$ case
and halo is long-lived against the gravitaional
radiation.

Another cooling mechanism, evaporation and collapse procedure,
is also inefficient\cite{sin1}.

In conclusion, we find that
 self-interactions between the particles, even weak,
  may play important role
in the boson star model of halos.

 Our work can be easily extended to the Boson-Fermion star\cite{fermion}
 and Q-star\cite{qstar}.

\section*{ ACKNOWLEDGMENTS }
 This work was supported in part by KOSEF.
One of authors(Lee) are thankful to M. Gleiser, P. Jetzer and
A. Liddle for helpful comments.

\newpage

\section*{ Figure Captions}

{\bf Figure 1}\\
Rotation velocities as a function of rescaled
$x$
for the parameters $\Lambda=0$(thin line)
 and  $\Lambda=300$(thick line).
$\Omega$ is $0.9$.
The real values of the $x$ end are $80$ and $220$
respectively.

\vspace{2cm}

{\bf Figure 2}\\
Rotation velocity and $10\times \sigma$
 as a function of position $x$
for $n=9$ solution.
The parameters are $\Lambda=300$, $\Omega=0.9$,
 $B(0)= 0.780 $ and $ \sigma(0)=0.01$.

\newpage
\begin{figure}[bp]
\caption{}
\setlength{\unitlength}{0.240900pt}
\ifx\plotpoint\undefined\newsavebox{\plotpoint}\fi
\sbox{\plotpoint}{\rule[-0.175pt]{0.350pt}{0.350pt}}%


\end{figure}

\newpage
\pagebreak
\begin{figure}[bp]
\caption{}

\setlength{\unitlength}{0.240900pt}
\ifx\plotpoint\undefined\newsavebox{\plotpoint}\fi
\sbox{\plotpoint}{\rule[-0.175pt]{0.350pt}{0.350pt}}%
%
\end{figure}

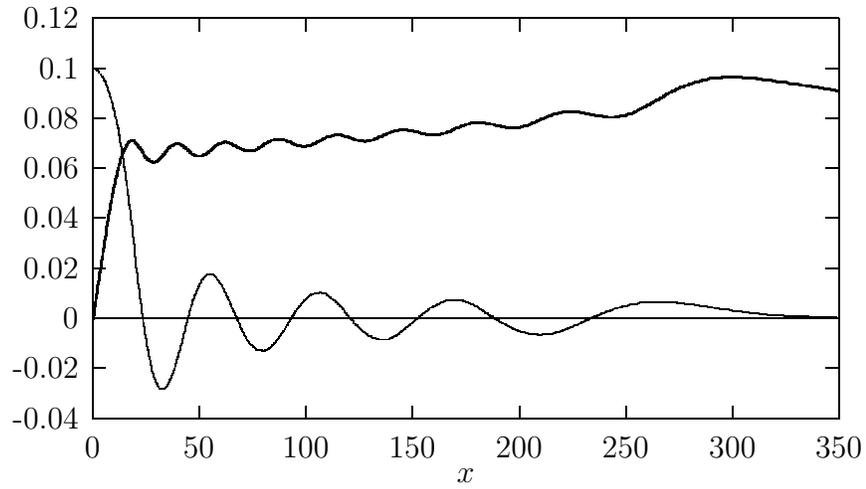
\end{document}